\documentclass[
aps,
prd,
reprint,
superscriptaddress,
floatfix,
nofootinbib,
longbibliography
]{revtex4-2}

\usepackage{graphicx}
\usepackage{amsmath,amssymb,amsfonts}
\usepackage{dcolumn}
\usepackage{bm}
\usepackage{mathtools}
\usepackage[caption=false]{subfig}
\usepackage[colorlinks,linkcolor=blue,anchorcolor=violet,citecolor=red,urlcolor=blue]{hyperref}
\usepackage{xcolor}
\usepackage{mathrsfs}
\usepackage{orcidlink}

\begin{document}

\title{Transient Information in Moving-Mirror Quantum Correlations}

\author{M.~Baran \"Okten \orcidlink{0000-0002-3484-9788}}
\email{okten26@itu.edu.tr}
\affiliation{Department of Physics Engineering, Faculty of Science and Letters, Istanbul Technical University, 34469 Maslak, Istanbul, T\"urkiye}

\author{Asl{\i} S.~Turan \orcidlink{0000-0003-0470-1552}}
\email{turanas26@itu.edu.tr}
\email{asli.turan@bogazici.edu.tr}
\affiliation{Department of Physics Engineering, Faculty of Science and Letters, Istanbul Technical University, 34469 Maslak, Istanbul, T\"urkiye}
\affiliation{Department of Physics, Bo\u{g}azi\c{c}i University, 34342 Bebek, Istanbul, T\"urkiye}

\author{\mbox{Michael~R.~R.~Good\,\orcidlink{0000-0002-0460-1941}}}
\email{muon@asu.edu}
\affiliation{Leung Center for Cosmology and Particle Astrophysics, National Taiwan University, Taipei 10617, Taiwan (R.O.C.)}
\affiliation{Physics Department \& Energetic Cosmos Laboratory, Nazarbayev University, Astana 010000, Qazaqstan}
\affiliation{Beyond Center for Fundamental Concepts in Science, Arizona State University, Tempe, Arizona 85287, USA}

\author{Kuan-Nan Lin \orcidlink{0000-0001-6561-4255}}
\email{knlinphy@gmail.com}
\affiliation{Asia Pacific Center for Theoretical Physics (APCTP), Pohang 37673, Korea}

\author{Pisin Chen \orcidlink{0000-0001-5251-7210}}
\email{pisinchen@phys.ntu.edu.tw}
\affiliation{Leung Center for Cosmology and Particle Astrophysics, National Taiwan University, Taipei 10617, Taiwan (R.O.C.)}
\affiliation{Department of Physics and Center for Theoretical Sciences, National Taiwan University, Taipei 10617, Taiwan (R.O.C.)}
\affiliation{Graduate Institute of Astrophysics, National Taiwan University, Taipei 10617, Taiwan (R.O.C.)}
\affiliation{Kavli Institute for Particle Astrophysics and Cosmology, SLAC National Accelerator Laboratory, Stanford University, Stanford, California 94305, USA}


\begin{abstract}
We study how a finite dynamical history remains encoded in a quantum field after a moving mirror returns to a common thermal trajectory. Two compact deformations of the Carlitz--Willey ray map are related by reversal of a control parameter and share the same asymptotic null line and Hawking temperature. The pairing induces an exact even--odd grading of the peeling field and Bogoliubov transformation. The complex continuum covariance retains the transient sign at first order, while the leading monochromatic quadratic response is even under reversal. Finite-bandwidth wave packets convert this distinction into signed interference with the thermal reference covariance. In linear response, the signed squared correlation is quartic in bandwidth for the selected right-moving packets and cubic for matched Rindler/Milne packets. The relative cross-covariance phase has a linear bandwidth dependence, with its leading coefficient maximized in magnitude by the flat window among normalized real symmetric windows. Finite-amplitude corrections introduce an additional linear bandwidth term in the squared Rindler/Milne response. The distinguishing packets sample the transient history in retarded time, while both trajectories share the same asymptotic Hawking temperature.
\end{abstract}

\maketitle

\section{Introduction}

An exactly thermal particle spectrum fixes the occupation numbers of outgoing modes. Correlations between those modes contain further information about the quantum state. This distinction appears in the study of Hawking radiation and black-hole entropy \cite{Bekenstein1973,Hawking1975,Hawking1976}, in Page's analysis of evaporation \cite{Page1993Entropy,Page1993Information}, and in descriptions of unitary evolution and horizon correlations \cite{StephensHooftWhiting1994, SusskindThorlaciusUglum1993, AlmheiriMarolfPolchinskiSully2013}. Calculations based on generalized entropy and replica geometries have since given explicit semiclassical Page curves \cite{EngelhardtWall2015, AlmheiriEngelhardtMarolfMaxfield2019, Penington2020, AlmheiriHartmanMaldacenaShaghoulianTajdini2020}. For a moving mirror, the dependence of the outgoing correlations on the radiation history can be calculated from the ray map.

An accelerated reflecting boundary maps incoming null rays into outgoing ones and reproduces the redshift mechanism underlying Hawking particle creation \cite{DeWitt1975,FullingDavies1976,DaviesFulling1977,BirrellDavies1982}. Derivatives of the ray map determine the conformal energy flux, while the complete map determines the Bogoliubov transformation. Carlitz and Willey supplied an exactly thermal trajectory \cite{CarlitzWilley1987}. An exact correspondence between a moving-mirror trajectory and null-shell black-hole collapse has also been established at the level of the Bogoliubov coefficients \cite{GoodAndersonEvans2016}. Finite-duration and dynamical trajectories extend the framework to transient radiation \cite{Walker1985,HottaShinoYoshimura1994,FabbriNavarroSalas2005,Haro2005}, and wave-packet constructions resolve the emitted field simultaneously in time and frequency \cite{GoodAndersonEvans2013}. Wave packets have also been used to describe accelerated mirrors as Gaussian quantum channels \cite{GianfeliciMancini2017}.

Extensions to higher dimensions have examined relativistic semitransparent mirrors, finite-size effects, and perfectly reflecting boundaries \cite{LinChouChen2021,LinChen2024,LinChen2025}. Accelerating plasma mirrors have also been proposed as laboratory analogs of Hawking evaporation \cite{ChenMourou2017}.

Long quasithermal epochs can preserve global purity \cite{Wilczek1993}. Time-resolved calculations show that energy flux and particle production resolve different features of the same trajectory \cite{GoodOng2015}. Horizonless mirrors can sustain extended thermal emission with finite total energy \cite{GoodYelshibekovOng2017}, and quasithermal trajectories exhibit controlled entropy evolution through long radiative epochs \cite{GoodLinderWilczek2020}. Asymptotically drifting mirrors have also been used to construct finite-energy, information-preserving models of null-shell collapse \cite{GoodOngMyrzakulYelshibekov2019}. More recent analytic trajectories further illuminate the relationship between energy flux and particle production \cite{GoodLinder2026}. Dynamical treatments incorporating backreaction and quantum coherence likewise preserve information beyond a thermal occupation law \cite{ChungVerlinde1994}.

In two-dimensional conformal field theory, energy flux and entanglement entropy are linked through the conformal map \cite{BianchiSmerlak2014}. Moving mirrors have consequently become useful models for Page-curve dynamics \cite{AkalEtAl2021PRL,AkalEtAl2021CQG}. Direct calculations of the radiation density matrix give R\'enyi entropies, modular flow, and conditions for unitary mirror evolution \cite{Reyes2021}, while broader classifications connect late-time trajectories with characteristic flux and entropy profiles \cite{AkalEtAl2022}. Chen and Yeom examined the entropy evolution of moving mirrors in connection with the information loss problem \cite{ChenYeom2017}.

Accelerated mirrors generate particle-partner correlations throughout the field \cite{ObadiaParentani2003}, and stress-tensor correlators retain detailed information about the trajectory \cite{FordRoman2004}. Purification partners of Hawking-like modes can be reconstructed from field correlations \cite{HottaSchutzholdUnruh2015}, with general partner constructions relating the partner mode to the measured mode and the underlying trajectory \cite{TomitsukaYamaguchiHotta2020}. The particle and energy costs of entanglement with late-time vacuum fluctuations have also been examined explicitly \cite{Wald2019}. Subsequent work has separated partner information from energetic final bursts \cite{OsawaEtAl2024}, developed second-order coherence as an entanglement-sensitive diagnostic \cite{TomonagaNambu2024}, and connected localized detection with field-entanglement extraction \cite{OsawaNambuYoshimoto2026}. Recent calculations have examined the relation between purification, energy flux, and correlations of localised detector modes \cite{Agullo2026,NambuYoshimoto2026}. 

The Rindler/Milne formulation of Lin, Chen, Good, and Nambu gives a particularly useful representation of these correlations \cite{LinChenGoodNambu2026}. For an asymptotically null mirror, reflected Hawking-like radiation and the complementary future sector enter a common Bogoliubov description that exposes squeezing, self-correlations, and cross-sector correlations. Finite frequency resolution adds a further ingredient. A wave packet samples neighboring frequencies coherently, so the thermal reference covariance can interfere with a transient correction. Packet bandwidth then becomes a controlled resolution scale for the dynamical history encoded in the state.

A related question arises in direction-resolved transient Hawking channels. Compact antipodally odd deformations of a conformal peeling field can be created and restored while leaving an integrated response to the intermediate history \cite{OktenTuran2026}. Antipodal relations also enter 't Hooft's horizon description, where correlations between antipodal sectors participate in the mapping between ingoing and outgoing degrees of freedom \cite{Hooft2016}. The moving-mirror construction developed here reduces the same paired-history idea to two trajectories related by reversal of a compact control parameter. Both trajectories return to the same asymptotic null line and Hawking temperature, allowing the quantum encoding of the intermediate history to be isolated directly.

We consider two mirror trajectories that differ during a compact interval and coincide with the Carlitz--Willey solution on either side. For the selected right-moving packet pair, the signed squared covariance begins at fourth order in bandwidth in linear response. Matched Rindler/Milne packets give a cubic contribution to the squared cross covariance and a linear contribution to its relative phase. At finite deformation amplitude, an additional contribution linear in bandwidth enters the squared cross covariance. The coefficients depend on weighted moments of the transient ray-map deformation.

\section{Paired transient mirror trajectories}
\label{sec:geometry}

We use units $c=\hbar=k_{\mathrm B}=1$ and the Minkowski metric with signature $(-,+)$. For a real massless scalar field in $(1+1)$ dimensions, let $t$ and $x$ be Minkowski time and position, with null coordinates $u=t-x$ and $v=t+x$. A perfectly reflecting mirror is specified by the ray map $v=P(u)$. Primes denote differentiation with respect to each function's argument. Its worldline can be parametrized by
\begin{equation}
 t(u)=\frac{u+P(u)}{2},
 \qquad
 x(u)=\frac{P(u)-u}{2},
 \label{eq:trajectory}
\end{equation}
which gives $dx/dt=[P'(u)-1]/[P'(u)+1]$. A future-directed timelike trajectory therefore satisfies $P'(u)>0$.

Following Carlitz and Willey \cite{CarlitzWilley1987}, we take the exactly thermal reference map with inverse-time scale $\kappa>0$,
\begin{equation}
 P_0(u)=-\frac{1}{\kappa}e^{-\kappa u}.
 \label{eq:P0}
\end{equation}
It approaches the future null line $v=0$. Since $P'_0=e^{-\kappa u}$, the peeling field $\kappa_0=-\partial_u\ln P'_0$ equals the constant $\kappa$, corresponding to the Hawking temperature $T_{\mathrm H}=\kappa/(2\pi)$.

The paired histories are compact additive deformations with real, dimensionless amplitude $\epsilon$ and a real-valued smooth profile $f$,
\begin{equation}
 P_{\sigma}(u;\epsilon)
 =
 P_0(u)+\sigma\epsilon f(u),
 \qquad
 \sigma=\pm1,
 \qquad
 f\in C_0^\infty(\mathbb{R}).
 \label{eq:Ppm}
\end{equation}
Both trajectories coincide exactly with the Carlitz--Willey solution outside the support of $f$. Their past and future asymptotic ray maps, future null line, and thermal scale are therefore identical. For any channel quantity, subscripts $\mathrm S$ and $\mathrm A$ denote its half-sum and half-difference, respectively.

The two trajectories represent effective conformal channels associated with the antipodally related sectors of Ref.~\cite{OktenTuran2026}. In the moving-mirror model, channel exchange is implemented by reversing the deformation parameter. The outgoing derivative-field two-point function returns to its Carlitz--Willey value when both retarded-time arguments lie after the compact interval. The frequency packets considered below have temporal tails overlapping that interval.

\begin{figure*}[t]
\centering
\subfloat[Ray maps $P_\sigma(u)$ compared with the Carlitz--Willey reference.\label{fig:ray-map}]{%
\includegraphics[width=0.31\textwidth]{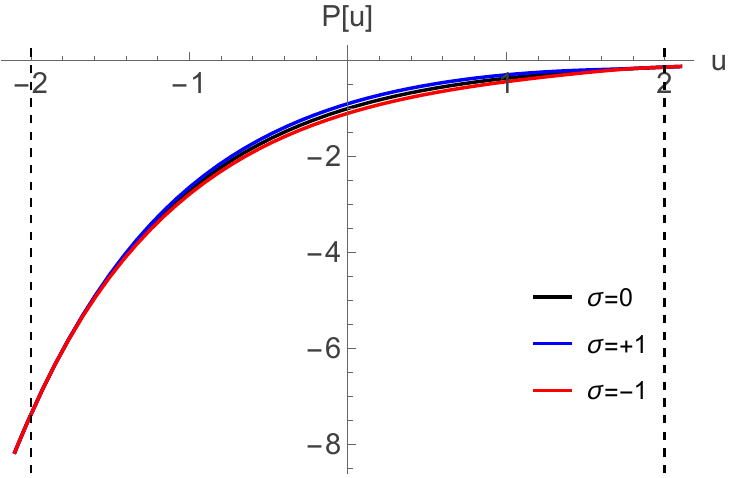}%
}
\hfill
\subfloat[Inverse ray maps $U_\sigma(v)$ compared with the common reference.\label{fig:inverse-ray-map}]{%
\includegraphics[width=0.31\textwidth]{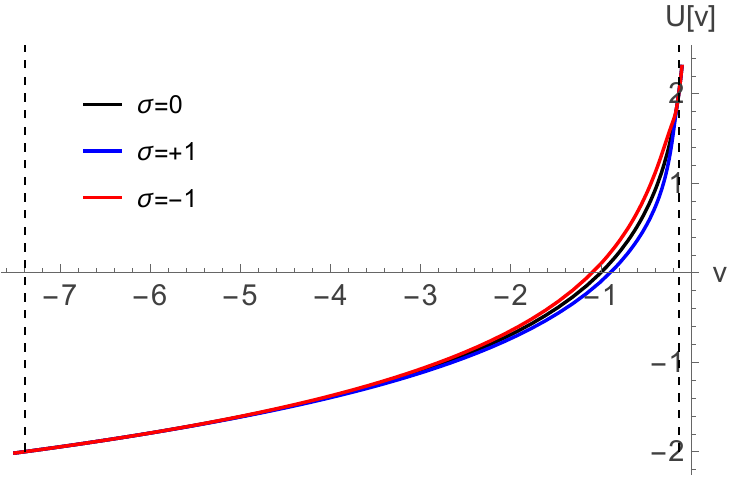}%
}
\hfill
\subfloat[Mirror trajectories $x(t)$ corresponding to the paired ray maps.\label{fig:mirror-trajectories}]{%
\includegraphics[width=0.31\textwidth]{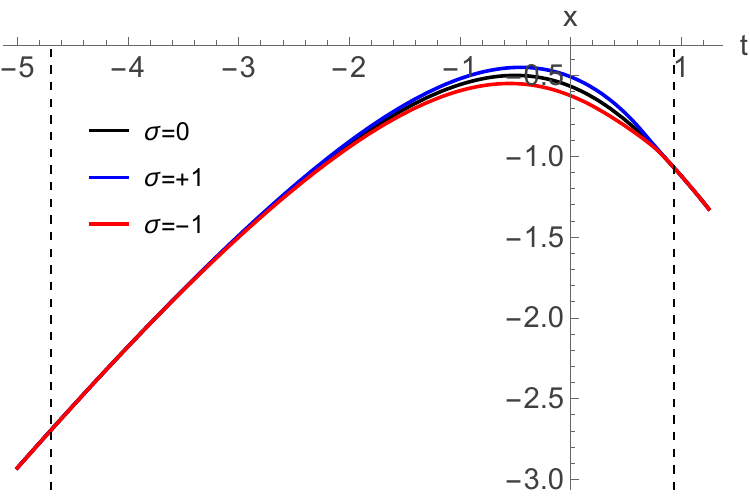}%
}
\caption{Compactly paired mirror histories for the fiducial deformation with $\kappa\tau=2$, $\kappa u_c=0$, and $\epsilon=0.10$. The ray map, inverse ray map, and spacetime trajectory encode the same finite disturbance. Outside the compact interval, the two channels coincide exactly with the Carlitz--Willey reference, while their intermediate histories are exchanged by $\epsilon\to-\epsilon$.}
\label{fig:paired-geometry}
\end{figure*}

Figure~\ref{fig:paired-geometry} shows the same compact history in three equivalent representations. Panel (a) compares the ray maps $P_\sigma(u)$ with the Carlitz--Willey reference, panel (b) compares the corresponding inverse ray maps $U_\sigma=P_\sigma^{-1}$ used in the Rindler/Milne construction in Sec.~\ref{sec:rindlermilne}, and panel (c) shows the mirror worldlines in Minkowski coordinates. The two channels separate only during the compact transient and recover the common reference history outside it.

For numerical illustrations, we use a smooth compact bump of half-duration $\tau$ centered at $u_c$. Its explicit form and timelike check are given in Appendix~\ref{app:technical}. The fiducial profile has $\kappa\tau=2$, $\kappa u_c=0$, and $\epsilon=0.10$.

Define
\begin{equation}
 r(u)=\frac{f'(u)}{P'_0(u)}=e^{\kappa u}f'(u).
 \label{eq:rdef}
\end{equation}
Then $P'_{\sigma}=P'_0(1+\sigma\epsilon r)$, and the pair remains timelike whenever $|\epsilon|\,\|r\|_\infty<1$, where $\|r\|_\infty=\sup_u|r(u)|$. Control-parameter reversal exchanges the trajectories exactly,
\begin{equation}
 P_\sigma(u;\epsilon)=P_{-\sigma}(u;-\epsilon).
 \label{eq:pairing}
\end{equation}

The corresponding peeling fields are
\begin{equation}
 \kappa_\sigma(u)
 =
 \kappa-\partial_u\ln\!\left(1+\sigma\epsilon r(u)\right).
 \label{eq:kappapm}
\end{equation}
Their channel-even and channel-odd combinations take the exact form
\begin{align}
 \kappa_{\mathrm S}
 &=
 \kappa-\frac{1}{2}\partial_u
 \ln\!\left(1-\epsilon^2r^2\right),
 &
 \kappa_{\mathrm A}
 &=
 -\partial_u\operatorname{artanh}(\epsilon r).
 \label{eq:kappaSA}
\end{align}
The geometry therefore carries an exact parity grading under reversal of the transient. At small amplitude,
\begin{align}
 \kappa_{\mathrm S}
 &=
 \kappa+\epsilon^2rr'+O(\epsilon^4),
 &
 \kappa_{\mathrm A}
 &=
 -\epsilon r'-\epsilon^3r^2r'+O(\epsilon^5).
 \label{eq:kappaSAexp}
\end{align}
The signed history enters the local peeling field at first order.

\begin{figure}[t]
\centering
\includegraphics[width=0.96\columnwidth]{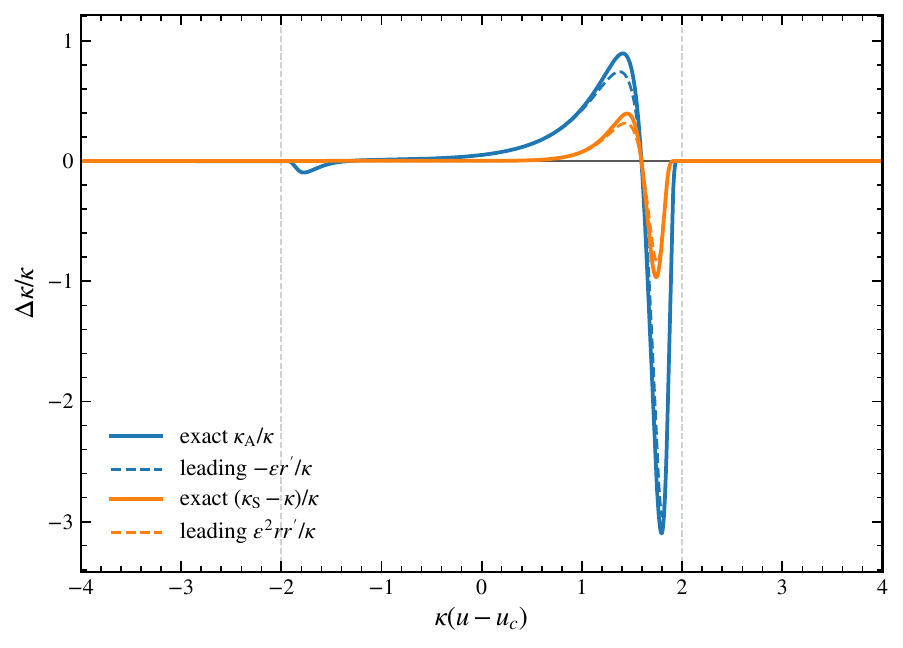}
\caption{Parity decomposition of the peeling response generated by the paired histories in Fig.~\ref{fig:paired-geometry}. Solid curves show the exact channel-even and channel-odd components, while dashed curves show their leading small-amplitude limits. Vertical lines mark the compact support of the transient.}
\label{fig:paired-peeling}
\end{figure}

The same grading appears in the radiated energy flux. Using the standard Schwarzian relation \cite{FullingDavies1976,DaviesFulling1977},
\begin{equation}
 F_\sigma(u)
 =
 -\frac{c_{\mathrm{CFT}}}{24\pi}\{P_\sigma,u\}
 =
 \frac{c_{\mathrm{CFT}}}{24\pi}
 \left(
 \dot\kappa_\sigma+\frac{1}{2}\kappa_\sigma^2
 \right),
 \label{eq:flux}
\end{equation}
where $c_{\mathrm{CFT}}$ is the conformal central charge ($c_{\mathrm{CFT}}=1$ for one real massless scalar), $\{P,u\}=P'''/P'-\tfrac32(P''/P')^2$ is the Schwarzian derivative, and an overdot denotes differentiation with respect to $u$. The channel-odd local flux begins as
\begin{equation}
 F_{\mathrm A}
 =
 -\frac{c_{\mathrm{CFT}}\epsilon}{24\pi}
 \left(r''+\kappa r'\right)
 +O(\epsilon^3).
 \label{eq:fluxA}
\end{equation}
Local energy transport therefore retains the transient sign linearly.

Integration over the restored history produces a different ordering. Define $\delta\kappa_\sigma=\kappa_\sigma-\kappa$ and the excess energy $\mathcal E_\sigma=\int_{-\infty}^{\infty}[F_\sigma(u)-F_0]\,du$ relative to the constant reference flux $F_0=c_{\mathrm{CFT}}\kappa^2/(48\pi)$. Compact restoration gives
\begin{equation}
 \mathcal E_\sigma
 =
 \frac{c_{\mathrm{CFT}}}{48\pi}
 \int_{-\infty}^{\infty}
 du\,\delta\kappa_\sigma^2(u).
 \label{eq:energyidentity}
\end{equation}
Its channel-odd component is
\begin{equation}
 \frac{\mathcal E_+-\mathcal E_-}{2}
 =
 -\frac{c_{\mathrm{CFT}}\epsilon^3}{24\pi}
 \int_{-\infty}^{\infty}
 du\,
 \frac{r(r')^2}{\left(1-\epsilon^2r^2\right)^2}.
 \label{eq:energyodd}
\end{equation}
The local peeling field and flux retain a linear signed response, while the integrated energetic distinction begins cubically. This separation anticipates the hierarchy in the quantum correlations, where frequency resolution determines the order at which the signed history enters a chosen observable.

\section{Bogoliubov response and right-moving correlations}
\label{sec:bogoliubov}

For the reflected right-moving sector we use the standard moving-mirror convention \cite{Walker1985,ObadiaParentani2001,GoodAndersonEvans2013,LinChenGoodNambu2026}. Here $\omega>0$ and $\omega'>0$ label outgoing and incoming frequency magnitudes; the negative sign on the incoming momentum subscript specifies its propagation direction:
\begin{align}
 \alpha^R_{\omega,-\omega'}[P]
 &=
 -\frac{1}{2\pi}
 \sqrt{\frac{\omega}{\omega'}}
 \int_{-\infty}^{\infty}
 du\,
 e^{-i\omega u+i\omega'P(u)},
 \label{eq:alpha}
 \\
 \beta^R_{\omega,-\omega'}[P]
 &=
 \frac{1}{2\pi}
 \sqrt{\frac{\omega}{\omega'}}
 \int_{-\infty}^{\infty}
 du\,
 e^{-i\omega u-i\omega'P(u)}.
 \label{eq:beta}
\end{align}
Let $a_{\omega'}$ annihilate an incoming positive-frequency mode in the in vacuum, and let $b_i$ annihilate an outgoing mode labeled by $i$. We use
\begin{equation}
 b_i
 =
 \int_0^\infty d\omega'
 \left(
 \overline{\alpha_{i,-\omega'}}a_{\omega'}
 -
 \overline{\beta_{i,-\omega'}}a^\dagger_{\omega'}
 \right).
 \label{eq:annihilationConvention}
\end{equation}

The trajectory pairing passes directly to the quantum transformation,
\begin{equation}
 \alpha_+(\epsilon)=\alpha_-(-\epsilon),
 \qquad
 \beta_+(\epsilon)=\beta_-(-\epsilon).
 \label{eq:Bogpair}
\end{equation}
The exact sine--cosine decomposition and the associated canonical identities are collected in Appendix~\ref{app:technical}. At first order,
\begin{equation}
\begin{aligned}
 \beta_\sigma
 &=
 \beta^{(0)}+\sigma\epsilon\beta^{(1)}+O(\epsilon^2),
 \\
 \alpha_\sigma
 &=
 \alpha^{(0)}+\sigma\epsilon\alpha^{(1)}+O(\epsilon^2).
\end{aligned}
 \label{eq:linearexp}
\end{equation}
The first-order coefficients are
\begin{align}
 \beta^{(1)}_{\omega,-\omega'}
 &=
 -\frac{i}{2\pi}\sqrt{\omega\omega'}
 \int_{-\infty}^{\infty}
 du\,
 f(u)e^{-i\omega u-i\omega'P_0(u)},
 \label{eq:beta1}
 \\
 \alpha^{(1)}_{\omega,-\omega'}
 &=
 -\frac{i}{2\pi}\sqrt{\omega\omega'}
 \int_{-\infty}^{\infty}
 du\,
 f(u)e^{-i\omega u+i\omega'P_0(u)}.
 \label{eq:alpha1}
\end{align}
The compact deformation therefore enters through finite-domain oscillatory integrals around an analytic thermal reference.

For the Carlitz--Willey trajectory the standard result \cite{CarlitzWilley1987} is
\begin{equation}
 \left|\beta^{(0)}_{\omega,-\omega'}\right|^2
 =
 \frac{1}{2\pi\kappa\omega'}
 \frac{1}{e^{2\pi\omega/\kappa}-1}.
 \label{eq:CWspectrum}
\end{equation}

The linear response also retains the complete compact deformation. Assume that $\beta^{(1)}_{\omega,-\omega'}[f]$ is zero for every positive $\omega$ and $\omega'$. Compact support permits the limit $\omega'\to0^+$ in Eq.~\eqref{eq:beta1}, giving
\begin{equation}
 \lim_{\omega'\to0^+}
 \frac{\beta^{(1)}_{\omega,-\omega'}}{\sqrt{\omega'}}
 =
 -\frac{i\sqrt{\omega}}{2\pi}
 \int_{-\infty}^{\infty}
 du\,f(u)e^{-i\omega u}.
 \label{eq:injectivelimit}
\end{equation}
The right-hand side is the Fourier transform of the real compactly supported deformation. Its value on the positive axis determines the transform on the negative axis by reality, and compact support fixes the function uniquely. The continuum first-order Bogoliubov map is therefore injective on the deformation class of Eq.~\eqref{eq:Ppm}. Finite-frequency observables select projections of this complete response, which makes their bandwidth dependence physically meaningful.

For the in-vacuum, define the normal and anomalous covariance kernels
\begin{align}
 \mathcal N_{\omega_1\omega_2}
 &=
 \int_0^\infty d\omega'\,
 \beta_{\omega_1,-\omega'}
 \overline{\beta_{\omega_2,-\omega'}},
 \label{eq:N}
 \\
 \mathcal M_{\omega_1\omega_2}
 &=
 \int_0^\infty d\omega'\,
 \beta_{\omega_1,-\omega'}
 \alpha_{\omega_2,-\omega'}.
 \label{eq:M}
\end{align}
With the convention of Eq.~\eqref{eq:annihilationConvention}, $\mathcal N_{ij}=\langle b_i^\dagger b_j\rangle$ and $\mathcal M_{ij}=-\overline{\langle b_i b_j\rangle}$. They satisfy the same channel pairing as the Bogoliubov transformation.

For the frequency difference $\Omega\in\mathbb{R}$, introduce the weighted transform
\begin{equation}
 H(\Omega)
 =
 \int_{-\infty}^{\infty}
 du\,e^{\kappa u}f(u)e^{-i\Omega u}.
 \label{eq:Htransform}
\end{equation}
Set $n_{\mathrm B}(\omega)=[\exp(2\pi\omega/\kappa)-1]^{-1}$ for the thermal Bose occupation number. For the Carlitz--Willey reference, direct reduction gives the first channel-odd coefficient,
\begin{equation}
\begin{aligned}
 \mathcal N^{(1)}_{\mathrm A,\omega_1\omega_2}
 & =
 \frac{i\sqrt{\omega_1\omega_2}}{2\pi}
 \left[n_{\mathrm B}(\omega_2)-n_{\mathrm B}(\omega_1)\right]
 \\
 & \quad \times H(\omega_1-\omega_2).
\end{aligned}
 \label{eq:N1reduced}
\end{equation}
The superscript $(1)$ denotes the coefficient linear in $\epsilon$ in the channel-odd part of the normal covariance. The complex covariance therefore carries the transient sign linearly.

For distinct continuum frequencies, define the channel-dependent quadratic right-moving diagnostic \begin{equation}
 Q^{RR}_\sigma(\omega_1,\omega_2) = |\mathcal N_{\sigma,\omega_1\omega_2}|^2
 + |\mathcal M_{\sigma,\omega_1\omega_2}|^2,
 \qquad
 \omega_1\neq\omega_2, \quad \sigma=\pm1.
 \label{eq:QRR}
\end{equation}
The Carlitz--Willey reference has zero off-diagonal covariance, so the first nonzero amplitude is linear in $\epsilon$, and its leading squared contribution is quadratic. The signed part consequently satisfies
\begin{equation}
 \frac{Q^{RR}_+-Q^{RR}_-}{2}=O(\epsilon^3).
 \label{eq:QRRantisymmetric}
\end{equation}
The continuum complex covariance resolves the sign at first order, while its leading squared magnitude carries the channel-even component.

\subsection{Finite-bandwidth right-moving correlations}
\label{subsec:RRpackets}

Finite-bandwidth modes change the interference structure. For frequency-bin label $j$, integer time label $n$, bandwidth $\Delta\omega>0$, and monochromatic outgoing mode $\phi_\omega$, use the packet basis \cite{GoodAndersonEvans2013},
\begin{equation}
 \phi_{jn}
 =
 \frac{1}{\sqrt{\Delta\omega}}
 \int_{j\Delta\omega}^{(j+1)\Delta\omega}
 d\omega\,
 e^{2\pi i\omega n/\Delta\omega}\phi_\omega.
 \label{eq:packet}
\end{equation}
The same transform applies to $\alpha$ and $\beta$.

We use $j\geq1$ and $I_j=[j\Delta\omega,(j+1)\Delta\omega]$, with central frequency $\bar\omega=(j+\tfrac12)\Delta\omega>\Delta\omega/2$. The narrow-band limit holds $\bar\omega$ fixed by taking narrower admissible bins successively with increasing $j$.

Let $C_\sigma$ denote the normal covariance between packets $(j,p)$ and $(j,0)$, with integer separation $p\geq1$ in the packet time label. Its channel expansion is
\begin{equation}
 C_\sigma
 =
 C_0+\sigma\epsilon C_1+\epsilon^2C_2+O(\epsilon^3),
 \label{eq:CpacketExpansion}
\end{equation}
and the signed squared correlation is
\begin{equation}
 |C_+|^2-|C_-|^2
 =
 4\epsilon\,\operatorname{Re}\!\left(\overline{C_0}C_1\right)
 +O(\epsilon^3).
 \label{eq:packetSignedInterference}
\end{equation}
Here $C_0$ is the thermal packet covariance, while $C_1$ and $C_2$ are its first- and second-order deformation coefficients. The thermal term supplies the reference amplitude for signed interference.

For integer separation $p\geq1$ and fixed $\bar\omega$, the narrow-band expansion gives
\begin{align}
 C_0^{(p)}
 &=
 -\frac{i\,n_{\mathrm B}'(\bar\omega)}{2\pi p}\Delta\omega
 +O\!\left((\Delta\omega)^2\right),
 \label{eq:packetC0Narrow}
 \\
 C_1^{(p)}
 &=
 -\frac{\bar\omega\,n_{\mathrm B}'(\bar\omega)H(0)}{4\pi^2p}
 (\Delta\omega)^2
 +O\!\left((\Delta\omega)^3\right).
 \label{eq:packetC1Narrow}
\end{align}
For a real deformation, the leading thermal term is imaginary and the leading transient term is real. Their quadrature relation removes the cubic interference. The first signed contribution is
\begin{equation}
\begin{aligned}
 &\left.
 \frac{\partial}{\partial\epsilon}
 \left(
 |C_+^{(p)}|^2-|C_-^{(p)}|^2
 \right)
 \right|_{\epsilon=0}
 \\
 &\qquad =
 \frac{H(0)[n_{\mathrm B}'(\bar\omega)]^2}{4\pi^4p^3}
 (\Delta\omega)^4
 +O\!\left((\Delta\omega)^5\right).
\end{aligned}
 \label{eq:packetSignedQuartic}
\end{equation}
For $H(0)\neq0$, this packet family has a leading signed squared correlation proportional to $(\Delta\omega)^4$ at fixed central frequency and fixed integer $p$. Appendix~\ref{app:packets} gives the intermediate packet expansion, Gaussian covariance bookkeeping, and finite-amplitude checks.

\section{Rindler/Milne correlations}
\label{sec:rindlermilne}

The right-moving sector describes the reflected Hawking-like radiation. For an asymptotically null mirror, the future Cauchy data also contain a complementary sector associated with the field beyond the limiting null ray. The Rindler/Milne construction gives a common mode description of the two sectors \cite{Unruh1976,LinChenGoodNambu2026}. We label the reflected sector by $R$ and the complementary Rindler sector by $L$; $\omega_R,\omega_L>0$ denote their frequency magnitudes. The compact deformation leaves the future null asymptote at $v=0$ fixed, so both channels share the same asymptotic partition. The inverse-map deformation used below is shown directly in Fig.~\ref{fig:inverse-ray-map}.

Let $U_\sigma=P_\sigma^{-1}$. For the reference map, $U_0(v)=-\kappa^{-1}\ln(-\kappa v)$ on $v<0$. Expanding the inverse relation $P_\sigma(U_\sigma(v))=v$ gives
\begin{equation}
 U_\sigma(v)
 =
 U_0(v)
 -
 \sigma\epsilon
 \frac{f(U_0(v))}{P'_0(U_0(v))}
 +O(\epsilon^2).
 \label{eq:U1}
\end{equation}

The reflected Minkowski out modes on future right null infinity $\mathscr I_R^+$ propagate backward into the $v<0$ part of past right null infinity $\mathscr I_R^-$. Their Bogoliubov coefficients are
\begin{align}
 \alpha_{\omega,-\omega'}^{R,\sigma}
 &= -\frac{1}{2\pi} \sqrt{\frac{\omega'}{\omega}} \int_{-\infty}^{0} dv\, e^{i\omega'v}e^{-i\omega U_\sigma(v)},
 \label{eq:bogoRalpha}
 \\
 \beta_{\omega,-\omega'}^{R,\sigma}
 & = -\frac{1}{2\pi} \sqrt{\frac{\omega'}{\omega}} \int_{-\infty}^{0} dv\, e^{-i\omega'v}e^{-i\omega U_\sigma(v)}.
\label{eq:bogoRbeta}
\end{align}
The complementary out modes occupy $v>0$ and are Rindler modes. Their explicit Gamma-function coefficients are given in Appendix~\ref{app:RL}; they depend on the common asymptotic partition, and the channel dependence of the cross correlations is carried by the reflected sector.

For positive frequencies, define
\begin{align}
 \mathcal N^{RL}_\sigma(\omega_R,\omega_L)
 &=
 \int_0^\infty d\omega'\,
 \beta^{R,\sigma}_{\omega_R,-\omega'}
 \overline{\beta^L_{-\omega_L,-\omega'}},
 \label{eq:NRL}
 \\
 \mathcal M^{RL}_\sigma(\omega_R,\omega_L)
 &=
 \int_0^\infty d\omega'\,
 \beta^{R,\sigma}_{\omega_R,-\omega'}
 \alpha^L_{-\omega_L,-\omega'}.
 \label{eq:MRL}
\end{align}
At the Carlitz--Willey point,
\begin{equation}
\begin{aligned}
 \mathcal N^{RL,(0)}(\omega_R,\omega_L)
 &=0,
 \\
 \mathcal M^{RL,(0)}(\omega_R,\omega_L)
 &=-\frac{\delta(\omega_R-\omega_L)}
 {2\sinh(\pi\omega_R/\kappa)}.
\end{aligned}
 \label{eq:RLCov0}
\end{equation}
Here $\delta$ denotes the Dirac delta distribution. The thermal anomalous covariance is concentrated on matched frequencies.

Using Eq.~\eqref{eq:Htransform}, the first channel-odd anomalous response reduces to
\begin{equation}
 \mathcal M^{RL,(1)}(\omega_R,\omega_L)
 =
 -\frac{i\sqrt{\omega_R\omega_L}}
 {4\pi\sinh(\pi\omega_L/\kappa)}
 H(\omega_R-\omega_L).
 \label{eq:MRL1}
\end{equation}
For real $f$, $H(0)$ is real. On matched-frequency support, the thermal anomalous covariance is real, and the first-order transient correction is imaginary. Their linear interference consequently vanishes as a distribution,
\begin{equation}
2\operatorname{Re}\left[
\overline{\mathcal M^{RL,(0)}}
\mathcal M^{RL,(1)}
\right]=0.
\label{eq:QRLcontinuumzero}
\end{equation}
The squared covariance is evaluated for finite wave packets in the following section. Appendix~\ref{app:RL} gives the full distributional reduction and the cancellation in terms of the underlying Bogoliubov integrals.

\section{Finite-bandwidth cross correlations}
\label{sec:finiteRL}

Packetization converts the matched continuum covariance into a finite cross-sector observable. For right- and left-moving packets, the packetized kernels are
\begin{align}
 \mathcal N^{RL,(q)}_{j n_R;k n_L}
 &=
 \frac{1}{\Delta\omega}
 \int_{I_j}d\omega_R
 \int_{I_k}d\omega_L\,
 \notag\\
 &\quad \times
 e^{2\pi i(\omega_Rn_R-\omega_Ln_L)/\Delta\omega}
 \mathcal N^{RL,(q)}(\omega_R,\omega_L),
 \label{eq:packetNRL}
 \\
 \mathcal M^{RL,(q)}_{j n_R;k n_L}
 &=
 \frac{1}{\Delta\omega}
 \int_{I_j}d\omega_R
 \int_{I_k}d\omega_L\,
 \notag\\
 &\quad \times
 e^{2\pi i(\omega_Rn_R+\omega_Ln_L)/\Delta\omega}
 \mathcal M^{RL,(q)}(\omega_R,\omega_L).
 \label{eq:packetMRL}
\end{align}
where $I_j$ is the frequency bin defined in Sec.~\ref{subsec:RRpackets}, the integer labels $n_R,n_L$ specify packet times, and $q=0,1$ denotes the thermal and first-order coefficients. The label $k$ indexes a second frequency bin.

For a pair of packet modes, let $\bm C^{RL}_\sigma$ denote the $2\times2$ quadrature cross-covariance block and $\|\cdot\|_{\mathrm F}$ the Frobenius matrix norm. Define
\begin{equation}
 Q^{RL}_\sigma
 =
\frac12\|\bm C^{RL}_\sigma\|_{\mathrm F}^2
 =
 |\mathcal N^{RL}_\sigma|^2
 +
 |\mathcal M^{RL}_\sigma|^2.
 \label{eq:QRLpacket}
\end{equation}
The channel-odd response has the form
\begin{equation}
\begin{aligned}
 \frac{Q^{RL}_+-Q^{RL}_-}{2}
 &=
 \epsilon\mathcal W_{RL}+O(\epsilon^3),
 \\
 \mathcal W_{RL}
 &=
 2\operatorname{Re}
 \!\left[
 \overline{\mathcal M^{RL,(0)}}\mathcal M^{RL,(1)}
 \right].
\end{aligned}
 \label{eq:WRLdef}
\end{equation}

Consider directly matched flat-window packets with $n_R=n_L=0$ and fixed central frequency. Define the weighted first moment
\begin{equation}
 H_1
 =
 \int_{-\infty}^{\infty}
 du\,u e^{\kappa u}f(u).
 \label{eq:H1}
\end{equation}
The leading transient packet covariance remains in quadrature with the thermal reference. Frequency averaging generates the first real interference at cubic order,
\begin{equation}
 \mathcal W_{RL}
 =
 \frac{\bar\omega H_1}{48\kappa}
 \frac{\cosh(\pi\bar\omega/\kappa)}
 {\sinh^3(\pi\bar\omega/\kappa)}
 (\Delta\omega)^3
 +
 O\!\left((\Delta\omega)^5\right).
 \label{eq:WRLcubic}
\end{equation}
Equation~\eqref{eq:WRLcubic} gives the linear-response coefficient in $\epsilon$. At finite deformation amplitude, the signed squared magnitude also contains a term of order $\epsilon^3\Delta\omega$, examined in Appendix~\ref{app:RL}.

For the compact profile used in Fig.~\ref{fig:paired-peeling}, the numerical fixed-central-frequency sequence follows this cubic law and approaches its analytic coefficient as the bandwidth decreases.

\begin{figure}[t]
\centering
\includegraphics[width=0.96\columnwidth]{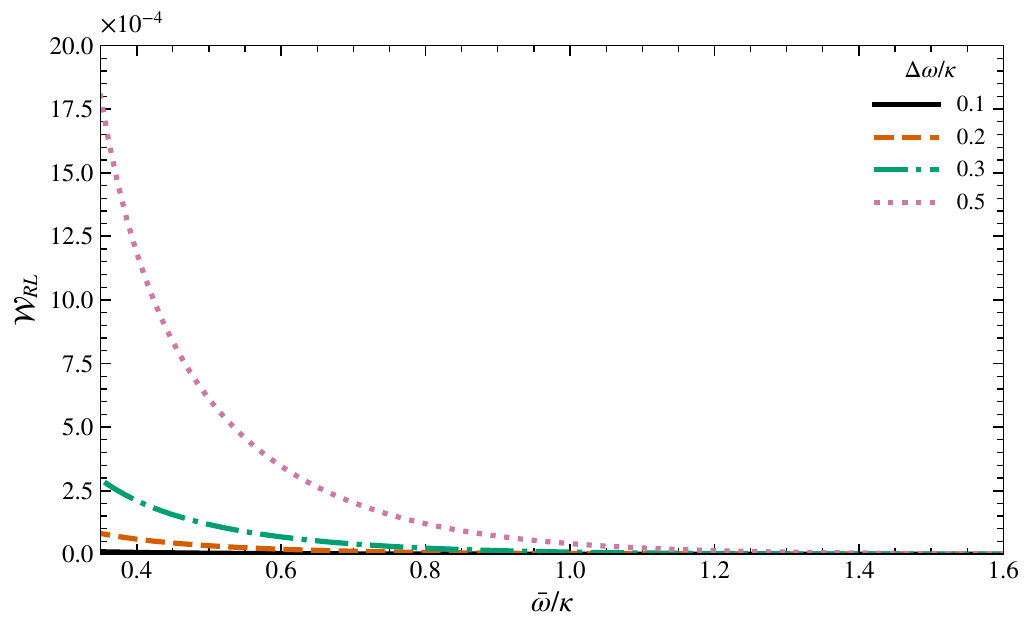}
\caption{Matched Rindler/Milne packet interference coefficient $\mathcal W_{RL}$ as a function of central frequency for several packet bandwidths. The curves approach the cubic narrow-band law of Eq.~\eqref{eq:WRLcubic}.}
\label{fig:RL-matched-response}
\end{figure}

\subsection{General packet windows}
\label{subsec:windows}

The cubic law also survives a change of frequency window. Let $s=(\omega-\bar\omega)/\Delta\omega$, and replace the flat frequency weight in Eq.~\eqref{eq:packet} by a real profile $g(s)$ supported on $[-1/2,1/2]$, with unit $L^2$ norm. The packet normalization factor remains $1/\sqrt{\Delta\omega}$. For moment order $\ell=0,1,2$, define
\begin{equation}
\begin{aligned}
\mu_\ell[g]
&=\int_{-1/2}^{1/2}ds\,s^\ell g(s),
\\
D[g]
&=\mu_0[g]\mu_2[g]-\mu_1[g]^2.
\end{aligned}
\label{eq:windowMoments}
\end{equation}
The narrow-band interference becomes
\begin{equation}
 \mathcal W_{RL}[g]
 =
 \frac{\bar\omega H_1}{4\kappa}
 \frac{\cosh(\pi\bar\omega/\kappa)}
 {\sinh^3(\pi\bar\omega/\kappa)}
 D[g](\Delta\omega)^3
 +
 O\!\left((\Delta\omega)^4\right).
 \label{eq:generalWRLcubic}
\end{equation}
For the flat window, $D[g]=1/12$, recovering Eq.~\eqref{eq:WRLcubic}. For nonnegative windows of finite width, $D[g]>0$, and the cubic response has a finite leading coefficient whenever $H_1\neq0$. The coefficient depends on the window moments. Signed windows can give $D[g]=0$, in which case higher-order terms control the response. The corresponding result for shifted packet time labels is given in Appendix~\ref{app:RL}.

\subsection{Phase-sensitive cross-covariance response}
\label{subsec:phaseRL}

The complex cross-covariance contains a lower-order signed response in its phase. For matched packets, define
\begin{equation}
 \Phi_\sigma
 =
 \arg\!\left[
 \overline{\mathcal M^{RL,(0)}}\mathcal M^{RL}_\sigma
 \right],
 \label{eq:PhiRL}
\end{equation}
with the branch chosen continuously from zero at $\epsilon=0$.

The channel-odd phase separation for flat-window packets is
\begin{equation}
 \frac{\Phi_+-\Phi_-}{2}
 =
 \epsilon\frac{\bar\omega H(0)}{2\pi}\Delta\omega
 +O\!\left(\epsilon(\Delta\omega)^3\right)
 +O\!\left(\epsilon^3\Delta\omega\right).
 \label{eq:PhiRLsigned}
\end{equation}

For a real symmetric normalized window,
\begin{equation}
\begin{aligned}
 \frac{\Phi_{+,g}-\Phi_{-,g}}{2}
 &=
 \epsilon\frac{\bar\omega H(0)}{2\pi}
 \mu_0[g]^2\Delta\omega
 \\
 &\quad
 +O\!\left(\epsilon(\Delta\omega)^3\right)
 +O\!\left(\epsilon^3\Delta\omega\right).
\end{aligned}
 \label{eq:generalPhiRLsigned}
\end{equation}
Cauchy--Schwarz gives $\mu_0[g]^2\leq1$. The flat window saturates this bound and therefore maximizes the leading phase response within the normalized real symmetric class. The $L^2$-normalized Hann window gives $\mu_0^2=2/3$, so its leading coefficient is two-thirds of the flat-window value.

For $H(0)\mu_0[g]\neq0$, the leading phase response is linear in bandwidth. The relative phase is invariant under common rephasings of the packet modes. Its measurement requires access to both sectors and a shared quadrature reference.

For the fiducial compact transient with $\epsilon=0.1$, $\Delta\omega=0.5\kappa$, and $\bar\omega=0.75\kappa$, direct evaluation with the full deformed ray maps gives a flat-window phase separation of $2.09^\circ$, compared with $2.08^\circ$ from the first-order calculation.

\subsection{Timing and amplitude of the compact transient}
\label{subsec:timing}

The leading phase coefficient determines $\epsilon H(0)$, and the cubic squared-magnitude coefficient determines $\epsilon H_1$. For a fixed bump shape and half-duration, their ratio gives the center relative to the retarded-time origin of the Carlitz--Willey reference, as shown in Eq.~\eqref{eq:timingRatio}. At the fiducial width, the weighted temporal offset from the bump center is $0.581119/\kappa$. Once the center is determined, the overall coefficient fixes the signed deformation amplitude within this prescribed profile family, given a calibrated thermal covariance and a shared quadrature reference. Finite-amplitude inference uses the full covariance, including the higher-order contribution in Appendix~\ref{app:RL}.

\section{Discussion and conclusion}

We considered two mirror histories with the same Carlitz--Willey asymptotes and opposite compact deformations. The complete first-order Bogoliubov response determines the deformation, while particular packet correlations retain selected features of it. In linear response, the signed squared correlation is quartic in bandwidth for the specified right-moving packet pair and cubic for matched Rindler/Milne packets. The relative cross-covariance phase has a linear bandwidth dependence. At finite amplitude, the squared cross-sector response also contains a cubic-in-amplitude contribution that is linear in bandwidth.

The weighted moments measured by the phase and squared-magnitude coefficients provide a further distinction between the timing and strength of the transient. With the bump profile and width fixed, their ratio determines its center relative to the reference retarded-time coordinate, and their scale determines its amplitude. This reconstruction assumes a common phase reference and linear-response calibration; finite-amplitude fits use the full covariance. Local derivative-field correlations recover the thermal reference once both observation times lie after restoration, while packets overlapping the transient retain its signature.

\begin{acknowledgments}
Funding comes partly from the FY2024-SGP-1-STMM Faculty Development Competitive Research Grant (FDCRGP) No.~201223FD8824 and SSH20224004 at Nazarbayev University in Qazaqstan. 

KNL was supported by an appointment as a YST fellow at the Asia Pacific Center for Theoretical Physics (APCTP) through the Korean Government's Science and Technology Promotion Fund and Lottery Fund.  

PC is supported by Taiwan's National Science and Technology Council (NSTC) under grant number 115-2112-M-002-014-, and by the Leung Center for Cosmology and Particle Astrophysics (LeCosPA), National Taiwan University.

The authors acknowledge COST Action CA23115 (Relativistic Quantum Information), funded by COST (European Cooperation in Science and Technology), for fostering research on quantum correlations in relativistic and gravitational settings.

Generative AI (OpenAI ChatGPT, GPT-6.1 Sol Ultra) was used for manuscript preparation and revision. The authors take full responsibility for the correctness and presentation of the content.

\end{acknowledgments}

\section*{Data Availability}

The data and code supporting the findings of this study are available from the authors upon reasonable request.

\appendix

\section{Geometry, flux identities, and Bogoliubov consistency}
\label{app:technical}

For completeness, the compact profile used in the numerical calculations is
\begin{equation}
\begin{aligned}
 f(u)
 &=
 \frac{1}{\kappa}
 B\!\left(\frac{u-u_c}{\tau}\right),
 \\
 B(y)
 &=
 \begin{cases}
 \displaystyle
 \exp\!\left(1-\frac{1}{1-y^2}\right), & |y|<1,\\[4pt]
 0, & |y|\geq1.
 \end{cases}
\end{aligned}
 \label{eq:bump}
\end{equation}
For this fixed bump profile, substitution of $y=(u-u_c)/\tau$ into the moments yields
\begin{equation}
\frac{H_1}{H(0)}
=u_c+\tau
\frac{\displaystyle\int_{-1}^{1}
dy\,y e^{\kappa\tau y}B(y)}
{\displaystyle\int_{-1}^{1}
dy\,e^{\kappa\tau y}B(y)}.
\label{eq:timingRatio}
\end{equation}
The profile $B(y)$ has unit peak at $y=0$ and support $|y|<1$. Hence $f$ is supported on $u_c-\tau<u<u_c+\tau$, and the peak departure of each ray map from $P_0$ is $\sigma\epsilon/\kappa$. The fiducial parameters are $\kappa\tau=2$, $\kappa u_c=0$, and $\epsilon=0.10$. For this profile, $|\epsilon|\|r\|_\infty=0.517$, comfortably inside the timelike domain.

The compact support of $f$ gives $r(\pm\infty)=0$. Writing $\delta\kappa_\sigma=\kappa_\sigma-\kappa$, Eq.~\eqref{eq:flux} gives the excess energy as
\begin{equation}
 \mathcal E_\sigma
 =
 \frac{c_{\mathrm{CFT}}}{24\pi}
 \int_{-\infty}^{\infty}
 du\,
 \left(
 \dot{\delta\kappa}_\sigma
 +\kappa\delta\kappa_\sigma
 +\frac12\delta\kappa_\sigma^2
 \right).
 \label{eq:Epreidentity}
\end{equation}
The first term is a boundary contribution. The second integrates to zero because $\delta\kappa_\sigma=-\partial_u\ln(1+\sigma\epsilon r)$ and the logarithm has equal asymptotic values. The remaining term gives Eq.~\eqref{eq:energyidentity}.

The exact peeling perturbation is
\begin{equation}
 \delta\kappa_\sigma
 =
 -\frac{\sigma\epsilon r'}{1+\sigma\epsilon r}.
 \label{eq:deltakappaExact}
\end{equation}
Substitution into Eq.~\eqref{eq:energyidentity}, followed by the channel-odd combination, gives Eq.~\eqref{eq:energyodd}.

The exact even--odd decomposition of the Bogoliubov transformation follows directly from Eq.~\eqref{eq:Ppm}:
\begin{align}
 \beta_{\mathrm S}
 &=
 \frac{1}{2\pi}\sqrt{\frac{\omega}{\omega'}}
 \int_{-\infty}^{\infty}
 du\,
 e^{-i\omega u-i\omega'P_0(u)}
 \cos\!\left(\epsilon\omega'f(u)\right),
 \label{eq:betaS}
 \\
 \beta_{\mathrm A}
 &=
 -\frac{i}{2\pi}\sqrt{\frac{\omega}{\omega'}}
 \int_{-\infty}^{\infty}
 du\,
 e^{-i\omega u-i\omega'P_0(u)}
 \sin\!\left(\epsilon\omega'f(u)\right).
 \label{eq:betaA}
\end{align}
For $\alpha$, the reference phase has the opposite sign. Its channel-even part retains the overall minus sign of Eq.~\eqref{eq:alpha}, while its channel-odd part has the same $-i$ prefactor as $\beta_{\mathrm A}$. Hence the $n$th perturbative coefficient acquires a factor $(-1)^n$ under channel exchange.

The first-order kernels satisfy the canonical row identities
\begin{align}
 \alpha^{(1)}\alpha^{(0)\dagger}
 +\alpha^{(0)}\alpha^{(1)\dagger}
 -\beta^{(1)}\beta^{(0)\dagger}
 -\beta^{(0)}\beta^{(1)\dagger}
 &=0,
 \label{eq:canonical1}
 \\
 \alpha^{(1)}\beta^{(0)T}
 +\alpha^{(0)}\beta^{(1)T}
 -\beta^{(1)}\alpha^{(0)T}
 -\beta^{(0)}\alpha^{(1)T}
 &=0.
 \label{eq:canonical2}
\end{align}
These relations provide analytic and numerical checks on the perturbative calculation.

Compact support also allows the thermal contribution to be separated before numerical integration:
\begin{align}
 \delta\beta_\sigma
 &=
 \frac{1}{2\pi}\sqrt{\frac{\omega}{\omega'}}
 \int_{\operatorname{supp}f}
 du\,
 e^{-i\omega u-i\omega'P_0(u)}
 \notag\\
 &\quad\times
 \left[
 e^{-i\sigma\epsilon\omega'f(u)}-1
 \right],
 \label{eq:compactSubBeta}
 \\
 \delta\alpha_\sigma
 &=
 -\frac{1}{2\pi}\sqrt{\frac{\omega}{\omega'}}
 \int_{\operatorname{supp}f}
 du\,
 e^{-i\omega u+i\omega'P_0(u)}
 \notag\\
 &\quad\times
 \left[
 e^{i\sigma\epsilon\omega'f(u)}-1
 \right].
 \label{eq:compactSubAlpha}
\end{align}
The finite-deformation evaluation can therefore be performed as a compact correction to the analytic Carlitz--Willey baseline.

\section{Right-moving packet asymptotics and finite covariance}
\label{app:packets}

For packet separation $p$, the Carlitz--Willey contribution may be written in the dimensionless frequency variable $s\in[0,1]$ as
\begin{equation}
 C_0^{(p)}
 =
 \int_0^1
 ds\,
 e^{2\pi i p s}
n_{\mathrm B}\!\left(\bar\omega+\left(s-\frac12\right)\Delta\omega\right). \label{eq:C0packetIntegral}
\end{equation}
Expanding the Planck factor about $\bar\omega$ gives
\begin{equation}
 C_0^{(p)}
 =
 n_{\mathrm B}'(\bar\omega)\Delta\omega
 \int_0^1
 ds\,
 \left(s-\frac12\right)e^{2\pi i p s}
 +O\!\left((\Delta\omega)^2\right),
 \label{eq:C0packetIntermediate}
\end{equation}
which yields Eq.~\eqref{eq:packetC0Narrow}.

Inserting Eq.~\eqref{eq:N1reduced} into the packet transform gives
\begin{equation}
\begin{aligned}
 C_1^{(p)}
 &=
 \frac{i}{2\pi\Delta\omega}
 \int_{I_j}d\omega_1
 \int_{I_j}d\omega_2\,
 e^{2\pi i p\omega_1/\Delta\omega}
 \\
 &\quad\times
 \sqrt{\omega_1\omega_2}
\left[n_{\mathrm B}(\omega_2)-n_{\mathrm B}(\omega_1)\right]
H(\omega_1-\omega_2).
\end{aligned}
 \label{eq:C1packetIntegral}
\end{equation}
Write $k_p=2\pi p$. All derivatives of $n_{\mathrm B}$ in the following coefficients are evaluated at $\bar\omega$. The leading terms appear in Eqs.~\eqref{eq:packetC0Narrow} and \eqref{eq:packetC1Narrow}. At the next orders, the coefficients of $(\Delta\omega)^2$ in $C_0^{(p)}$ and $(\Delta\omega)^3$ in $C_1^{(p)}$ are
\begin{align}
C_{0,2}
&=\frac{n_{\mathrm B}''(\bar\omega)}{k_p^2},
\\
C_{1,3}
&=-\frac{i}{2\pi k_p^2}
\left[
H(0)(n_{\mathrm B}'+\bar\omega n_{\mathrm B}'')
+2\bar\omega n_{\mathrm B}'H'(0)
\right].
\end{align}
For real $f$, $H'(0)=-iH_1$. The terms containing $n_{\mathrm B}''$ cancel in the real part of the thermal-transient interference. The remaining coefficient gives Eq.~\eqref{eq:packetSignedQuartic}.

For packet pairs $(j,n+p)$ and $(j,n)$ with $n(n+p)\neq0$, the corresponding leading signed response is $O(\epsilon(\Delta\omega)^5)$. The quartic coefficient in the main text applies to the family $(j,p)$ and $(j,0)$.

For a collection of $m$ packet modes, labelled by $a=1,\ldots,m$, the normal and anomalous kernels determine a finite Gaussian covariance matrix. Using the standard continuous-variable convention \cite{Weedbrook2012}, define
\begin{equation}
\begin{aligned}
 X_a&=\frac{b_a+b_a^\dagger}{\sqrt2},
 &
 Y_a&=\frac{b_a-b_a^\dagger}{i\sqrt2},
 \\
 \bm\xi&=(X_1,Y_1,\ldots,X_m,Y_m)^T.
\end{aligned}
 \label{eq:quadratures}
\end{equation}
The covariance matrix has entries $V_{ij}$, with $\Delta\xi_i=\xi_i-\langle\xi_i\rangle$:
\begin{equation}
 V_{ij}
 =
 \frac12
 \left\langle
 \Delta\xi_i\Delta\xi_j+\Delta\xi_j\Delta\xi_i
 \right\rangle.
 \label{eq:V}
\end{equation}
The exact channel pairing gives
\begin{equation}
 V_+(\epsilon)=V_-(-\epsilon),
 \qquad
 \frac{V_+-V_-}{2}
 =
 \epsilon V_{\mathrm A}^{(1)}+O(\epsilon^3).
 \label{eq:VA}
\end{equation}
Thus the finite covariance matrix carries a linear channel-odd component even when a particular quadratic contraction begins at a higher bandwidth power.

The finite-deformation calculation was evaluated independently using the full maps $P_\sigma=P_0+\sigma\epsilon f$. Extrapolation to $\epsilon\to0$ agrees with the reduced first-order kernel in Eq.~\eqref{eq:N1reduced}, and fixed-central-frequency sequences reproduce the coefficient in Eq.~\eqref{eq:packetSignedQuartic}.

\section{Rindler/Milne reduction, window dependence, and numerical checks}
\label{app:RL}

The complementary $v>0$ out modes are the standard Rindler modes \cite{Unruh1976,LinChenGoodNambu2026}. Their Bogoliubov coefficients are
\begin{align}
 \alpha^L_{-\omega,-\omega'}
 &=
 -\frac{i}{2\pi\omega'}
 \sqrt{\frac{\omega'}{\omega}}
 e^{\pi\omega/(2\kappa)}
 \left(\frac{\kappa}{\omega'}\right)^{-i\omega/\kappa}
 \Gamma\!\left(1-\frac{i\omega}{\kappa}\right),
 \label{eq:bogoLalpha}
 \\
 \beta^L_{-\omega,-\omega'}
 &=
 \frac{i}{2\pi\omega'}
 \sqrt{\frac{\omega'}{\omega}}
 e^{-\pi\omega/(2\kappa)}
 \left(\frac{\kappa}{\omega'}\right)^{-i\omega/\kappa}
 \Gamma\!\left(1-\frac{i\omega}{\kappa}\right).
 \label{eq:bogoLbeta}
\end{align}

The first-order anomalous cross covariance follows from Eqs.~\eqref{eq:beta1} and \eqref{eq:bogoLalpha}. Regulate the incoming frequency integral by $e^{-\eta\omega'}$, with $\eta>0$. The Mellin integral gives
\begin{equation}
\begin{aligned}
&\lim_{\eta\to0^+}
\int_0^\infty d\omega'\,
e^{-\eta\omega'}
\left(\frac{\kappa}{\omega'}\right)^{
-i\omega_L/\kappa}
e^{-i\omega'P_0(u)}
\\
&\quad =
i\kappa e^{\kappa u+i\omega_Lu}
e^{-\pi\omega_L/(2\kappa)}
\Gamma\left(1+\frac{i\omega_L}{\kappa}\right).
\end{aligned}
\end{equation}
Combining the Gamma functions through $\Gamma(1+ix)\Gamma(1-ix) =\pi x/\sinh(\pi x)$ leaves the weighted Fourier transform $H(\omega_R-\omega_L)$. This gives Eq.~\eqref{eq:MRL1}. On matched frequencies the first-order contribution is imaginary for real $f$, giving the distributional cancellation in Eq.~\eqref{eq:QRLcontinuumzero}.

For a general real packet window $g$, the cubic coefficient follows by expanding both frequency integrals around $\bar\omega$. The combination $D[g]$ in Eq.~\eqref{eq:windowMoments} collects the first two window moments entering the real thermal-transient interference. For nonnegative windows of finite width, $D[g]>0$, and the sign of the cubic coefficient follows the sign of $H_1$. Frequency-matched packet pairs with nonzero opposite time labels retain the cubic bandwidth power with an additional time-label suppression.

For the phase response, symmetry $g(-s)=g(s)$ removes the even bandwidth correction between the leading and cubic terms. The ratio of first-order transient to thermal covariance has the expansion
\begin{equation}
 \frac{\mathcal M^{RL,(1)}_g}{\mathcal M^{RL,(0)}_g}
 =
 i\,
 \frac{\bar\omega H(0)}{2\pi}
 \mu_0[g]^2\Delta\omega
 +O\!\left((\Delta\omega)^3\right).
 \label{eq:generalMratio}
\end{equation}
Its imaginary part yields Eq.~\eqref{eq:generalPhiRLsigned}. Normalization gives $\mu_0[g]^2\leq1$, with equality for the flat window.

For comparison, the normalized Hann window is
\begin{equation}
 g_{\mathrm H}(s)
 =
 \sqrt{\frac{2}{3}}
 \left[1+\cos(2\pi s)\right],
 \qquad
 |s|\leq\frac12,
 \label{eq:HannWindow}
\end{equation}
and has $\mu_0[g_{\mathrm H}]^2=2/3$.

For the fiducial transient, the fixed-central-frequency cubic sequence approaches the analytic coefficient predicted by Eq.~\eqref{eq:WRLcubic}. At $\bar\omega=0.75\kappa$, the limiting value of $\mathcal W_{RL}/(\Delta\omega)^3$ is $1.10519\times10^{-3}$ for $\kappa=1$. The weighted zeroth moment is $H(0)=3.26750$ in the same units.

Direct evaluation with the full deformed maps was also performed at finite amplitude. For $\epsilon=0.1$ and $\Delta\omega=0.5\kappa$, the flat-window phase separation is $2.09^\circ$, compared with $2.08^\circ$ from the first-order calculation. For the squared-magnitude response, the limits in deformation amplitude and bandwidth have distinct leading contributions. At $\kappa=1$ and $\bar\omega=0.75$, the fiducial profile gives
\begin{equation}
\begin{aligned}
\frac{Q^{RL}_+-Q^{RL}_-}{2}
={}&0.00110519\,\epsilon(\Delta\omega)^3\\
&-0.00587898\,\epsilon^3\Delta\omega+\cdots.
\end{aligned}
\end{equation}
At $\epsilon=0.1$ and $\Delta\omega=0.1$, the full normal and anomalous covariance calculation yields $(Q^{RL}_+-Q^{RL}_-)/(2\epsilon) =-4.64\times10^{-6}$. The first-order contribution at these parameters is $1.11\times10^{-6}$. The cubic bandwidth coefficient is recovered by taking the linear-response limit first.

\bibliography{main}

\end{document}